\documentstyle{article}
\begin{document}
\large
\baselineskip=24pt
\title{Lattice Boltzmann Thermohydrodynamics}
\vspace{1in}
\author{F. J. Alexander, S. Chen and J. D. Sterling\thanks{Permanent Address:
Advanced Projects Research Incorporated, 5301 N. Commerce Ave.,
Suite A, Moorpark, CA 93021}\\ Center for Nonlinear Studies and Theoretical
Division\\ Los Alamos National Laboratory\\
 Los Alamos, NM 87545}
\date{}
\maketitle
\vspace{0.4in}
\begin{center}
{\bf ABSTRACT}
\end{center}
We introduce a lattice Boltzmann computational scheme capable
of modeling thermohydrodynamic flows of monatomic gases.  The parallel
nature of this approach provides a numerically efficient alternative to
traditional methods of computational fluid dynamics.  The scheme uses
a small number of discrete velocity states and a linear,
single-time-relaxation collision operator. Numerical simulations
in two dimensions agree well with exact solutions for
adiabatic sound propagation and Couette flow with heat transfer.
\pagebreak

The lattice Boltzmann (LB) method is a discrete, in space and time,
microscopic, kinetic equation description for the evolution of the
velocity distribution function of a fluid~\cite{zanetti,hig,succi1}.
Like lattice gas (LG) automata~\cite{fhp}, LB methods are well suited
for simulating a variety of physical systems in a parallel computing
environment. As a result, the LB approach has found recent successes in a host
of fluid dynamical problems, including flows in porous media~\cite{guns},
magnetohydrodynamics~\cite{syc1}, immiscible fluids~\cite{guns} and
turbulence~\cite{succi,syc2}.  Its efficiency competes with,
and in some cases exceeds, that of traditional numerical methods,
while its physical interpretation is transparent.

Noticeably absent, though, from the list of successful applications
of LG and LB methods is a model which can simulate the full set
of thermohydrodynamic equations.
Previous attempts at developing such a model
have {\em exclusively} involved LG automata~\cite{bzk,syc3} whose
Fermi-Dirac equilibrium
distributions do not have sufficient flexibility to guarantee the correct form
of the energy equation (3).  LB methods are considerably more
flexible, but have not, until now, been applied to this problem.

The thermohydrodynamic equations of classical kinetic theory result
from a Chapman-Enskog expansion of the {\em continuum} Boltzmann
equation with the assumption of a  Maxwellian equilibrium distribution.
Since an exact Maxwellian distribution with a continuous
distribution of velocities, both in angle and magnitude, cannot be
implemented on a system that is discrete in both space
and time, we seek an alternative distribution which will nevertheless
give rise to the same macroscopic physics.
In this Letter we address this issue and introduce
a LB scheme which can simulate the following continuity,
momentum, and
energy equations for viscous, compressible, and heat-conducting flows:
\begin{eqnarray}
\frac{\partial n }{\partial t} + \frac{\partial}
{\partial x_{\alpha}}(nu_{\alpha}) = 0 &,
\end{eqnarray}
\pagebreak
\begin{eqnarray}
n\frac{\partial u_{\alpha}}{\partial t}+nu_{\beta}\frac{\partial
u_{\alpha}}{\partial x_{\beta}}=
-\frac{\partial p}{\partial x_{\alpha}}+\frac{\partial}
{\partial x_{\alpha}}(\lambda\frac{\partial u_{\gamma}}
{\partial x_{\gamma}})+\frac{\partial}
{\partial x_{\beta}}(\mu(\frac{\partial u_{\beta}}
{\partial x_{\alpha}}+\frac{\partial u_{\alpha}}
{\partial x_{\beta}}))
&  ,
\end{eqnarray}
and
\begin{eqnarray}
n\frac{\partial \epsilon}{\partial t}+
nu_{\alpha}\frac{\partial
\epsilon}{\partial x_{\alpha}}=-p\frac{\partial u_{\gamma}}
{\partial x_{\gamma}}+
\frac{\partial}
{\partial x_{\beta}}(\kappa\frac{\partial T}
{\partial x_{\beta}})+
\mu(\frac{\partial u_{\alpha}}
{\partial x_{\beta}}+\frac{\partial u_{\beta}}
{\partial x_{\alpha}})\frac{\partial u_{\beta}}
{\partial x_{\alpha}}+\lambda
(\frac{\partial u_{\gamma}}{\partial x_{\gamma}})^{2}
&,
\end{eqnarray}
where $n$ is the fluid mass density, $\epsilon$ is the internal energy
per unit mass and is proportional to the temperature
$T$,
${\bf u}$ is the local velocity, $p$ is the pressure,
and $\lambda$, $\mu$, and $\kappa$
are the second viscosity,
shear viscosity, and thermal conductivity, respectively.

The starting point of the LB method is the kinetic equation
for the velocity distribution function, $f_{\sigma i} ({\bf x}, t)$:
\begin{equation}
f_{\sigma i} ({\bf x} + {\bf e}_{\sigma i }, t+1) -
f_{\sigma i} ({\bf x},t) =
\Omega_{\sigma i},
\end{equation}
where the nonnegative, real number $f_{\sigma i} ({\bf x},t)$
is the mass of fluid
at lattice node ${\bf x}$ and time $t$, moving in direction $i$
with speed,  $ |{\bf e}_{\sigma i }| = \sigma$, $\sigma =1,2,...N$, where
$N$ is the number of speeds.
The $\sigma = 0$ speed corresponds to the component of the fluid which
is at rest. The term $\Omega_{\sigma i}$ represents
the rate of change of $f_{\sigma i}$ due to collisions.
For computational efficiency, it is desirable to
find the minimal set of $\sigma$ and $i$, for which a coarse-graining
of the kinetic equation (4) leads to the macroscopic dynamics of interest.

The microscopic dynamics associated with Equation (4) can be viewed as
a two step process: free streaming and collision.  During the free streaming
step, $f_{\sigma i}({\bf x}+{\bf e}_{\sigma i})$ is replaced by
$f_{\sigma i}(\bf{x})$.  Thus, each site exchanges mass with its
neighbors, i.e. sites
connected by lattice vectors ${\bf e}_{\sigma i}$.
In the collision step the distribution functions at each site
then relax toward a state of local equilibrium.
For simplicity, we consider the linearized, single-time-relaxation model
of Bhatnagar, Gross, and Krook~\cite{BGK},
which has recently been applied to LB models~\cite{syc1,qian,syc4,koelman}:
\begin{equation}
\Omega_{\sigma i}= -\frac{1}{\tau} (f_{\sigma i}-f_{\sigma i}^{eq}).
\end{equation}
The collision operator $\Omega_{\sigma i}$  conserves the local mass,
momentum and kinetic energy: $ \sum_{\sigma i} \Omega_{\sigma i} = 0,
\sum_{\sigma i} \Omega_{\sigma i}{\bf e}_{\sigma i } = 0,$ and
$ \sum_{\sigma i} \Omega_{\sigma i}{\bf e}_{\sigma i}^{2}/2 = 0$,
and the parameter, $\tau$, controls the rate at which the system relaxes to the
local equilibrium, $f_{\sigma i}^{eq}$.

The LB method, unlike LGs, has considerable flexibility
in the choice of the local equilibrium distribution.
A general equilibrium distribution is given by a truncated power series
in the local velocity $\bf{u}$, valid for $|{\bf  u}| \ll 1$,
\begin{equation}
f_{\sigma i}^{\rm eq} = A_{\sigma} + B_{\sigma}{\bf e}_{\sigma i}\cdot
{\bf u}
+C_{\sigma}({\bf e}_{\sigma i}\cdot{\bf  u})^{2}
+D_{\sigma}u^{2}
+E_{\sigma}({\bf e}_{\sigma i}\cdot{\bf  u})^{3}
+F_{\sigma}({\bf e}_{\sigma i}\cdot{\bf  u})u^{2},
\end{equation}
where the velocity is defined by: $n {\bf u} =
\sum_{\sigma i}f_{\sigma i}{\bf e}_{\sigma i}$.
The coefficients, $A,B,...,F$, are functions of the local
density $n = \sum_{\sigma i}f_{\sigma i}$ and internal energy
$n \epsilon = \sum_{\sigma i}f_{\sigma i}({\bf e}_{\sigma i}-{\bf u})^2/2$,
and their functional forms depend on the geometry of the underlying
lattice.

The long-wavelength, low-frequency behavior of the this system
is obtained by a Taylor series expansion of Equation (4)
to second order in the lattice spacing and time step:
\begin{eqnarray}
\frac{\partial f_{\sigma i}}{\partial t}
+{\bf e}_{\sigma i} \cdot {\bf \nabla}f_{\sigma i}
+\frac{1}{2} {\bf e}_{\sigma i} {\bf e}_{\sigma i} :{\bf \nabla}
{\bf \nabla} f_{\sigma i}
+{\bf e}_{\sigma i}
\cdot {\bf \nabla}\frac{\partial}{\partial t}f_{\sigma i}
+\frac{1}{2}\frac{\partial}{\partial t}\frac{\partial}{\partial t}
f_{\sigma i} = \Omega_{\sigma i}.
\end{eqnarray}
In order to derive the macroscopic hydrodynamic equations,
we adopt the following Chapman-Enskog multi-scale expansions.
We expand the time derivative as
\begin{equation}
\frac{\partial}{\partial t} =
\varepsilon \frac{\partial}{\partial t_{1}}+
\varepsilon^{2} \frac{\partial}{\partial t_{2}} + ...,
\end{equation}
where the lower order terms in $\varepsilon$ vary more rapidly.
Because we are interested in small departures from local equilibrium,
we expand the distribution function as
\begin{equation}
f_{\sigma i} = f_{\sigma i}^{eq}
+\varepsilon f_{\sigma i}^{(1)}  +\varepsilon^{2}f_{\sigma i}^{(2)} + ...,
\end{equation}
and the collision operator as
\begin{equation}
\frac{\Omega_{\sigma i}}{\varepsilon} =
-\frac{1}{\tau \varepsilon}
(\varepsilon f_{\sigma i}^{(1)}  +\varepsilon^{2}f_{\sigma i}^{(2)} + ...).
\end{equation}

Substituting the above expansions into the kinetic equation, we find
\begin{eqnarray}
\frac{\partial}{\partial t_{1}}f^{eq}_{\sigma i}
+{\bf e}_{\sigma i} \cdot {\bf \nabla} f^{eq}_{\sigma i} = -\frac{1}{\tau}
f^{(1)}_{\sigma i}
\end{eqnarray}
to order $\varepsilon$,  and
\begin{eqnarray}
\frac{\partial}{\partial t_{1}}f^{(1)}_{\sigma i}+
\frac{\partial}{\partial t_{2}}f^{eq}_{\sigma i}
+{\bf e}_{\sigma i} \cdot {\bf \nabla} f^{(1)}_{\sigma i}
+\frac{1}{2} {\bf e}_{\sigma i} {\bf e}_{\sigma i} :{\bf \nabla}
{\bf \nabla} f^{eq}_{\sigma i}+
{\bf e}_{\sigma i} \cdot {\bf \nabla} \frac{\partial}{\partial t_{1}}
f^{eq}_{\sigma i}+ \frac{1}{2} \frac{\partial^{2}}{\partial t_{1}^{2}}
f^{eq}_{\sigma i}
=-\frac{1}{\tau}
f^{(2)}_{\sigma i}
\end{eqnarray}
to order $\varepsilon^{2}$.
With Equation (11) and some algebra, we can rewrite Equation (12)
as:
\begin{eqnarray}
\frac{\partial}{\partial t_{2}}f^{eq}_{\sigma i}
+(1-\frac{1}{2\tau})(\frac{\partial}{\partial t_{1}}f^{(1)}_{\sigma i}
+{\bf e}_{\sigma i} \cdot {\bf \nabla} f^{(1)}_{\sigma i}) =-\frac{1}{\tau}
f^{(2)}_{\sigma i}.
\end{eqnarray}

Summing moments of Equations (11) and (13), we obtain to order
$\varepsilon^2$, the continuity equation,
\begin{equation}
\frac{\partial n}{\partial t} +\nabla \cdot n{\bf u} = 0,
\end{equation}
the momemtum equation,
\begin{equation}
\frac{\partial n{\bf u}}{\partial t} + \nabla \cdot {\bf \Pi}  = 0,
\end{equation}
and the energy equation,
\begin{equation}
\frac{\partial n\epsilon}{\partial t} + \nabla\cdot {(n\epsilon{\bf u})} +
\nabla\cdot{\bf q} + {\bf P}:\nabla {\bf u} = 0.
\end{equation}
The momentum flux tensor ${\bf \Pi}
= \sum_{\sigma i}
[f^{eq}_{\sigma i} + (1 - \frac{1}{2\tau})f^{(1)}_{\sigma i}]
{\bf e}_{\sigma i}
{\bf e}_{\sigma i}$; the heat flux,
${\bf q}_{\alpha} = (1/2)\sum_{\sigma i}
[f^{eq}_{\sigma i} + (1 - \frac{1}{2\tau})f^{(1)}_{\sigma i}]
({\bf e}_{\sigma i} - {\bf u})^2
({\bf e}_{\sigma i}- {\bf u})_{\alpha}$, and
$\bf P$ is the pressure tensor,
${\bf P}_{\alpha\beta} = (1/2)\sum_{\sigma i}[f^{eq}_{\sigma i} + (1 -
\frac{1}{2\tau})f^{(1)}_{\sigma i}]
({\bf  e}_{\sigma i} - {\bf u})_{\alpha}
({\bf e}_{\sigma i}- {\bf u})_{\beta}$.

To recover the Euler equations, we neglect the order $\varepsilon^2$ terms
and impose four further constraints on the equilibrium distribution
function.
The first of these constraints requires that the momentum flux tensor,
${\bf \Pi}_{\alpha \beta}^{eq}$, be isotropic.  The velocity independent
portion
of the tensor is then identified as the pressure, and this immediately
results in the equation of state for an ideal gas, $p=n\epsilon$.
The remaining two constraints require that the convective terms be
Galilean invariant, and that the heat flux vanish
to first order in $\varepsilon$, ${\bf q}^{(eq)}=0$.
Thus we obtain the equations for compressible, inviscid and
nonconducting flow of a monatomic gas.

Retaining terms to order $\varepsilon^2$ and imposing two additional
constraints, we recover the Navier-Stokes level equations. These constraints
are that the momentum flux tensor, ${\bf \Pi}^{(1)}$,
be isotropic and that the heat flux, ${\bf q}^{(1)}$, be
proportional to the gradient of the temperature: ${\bf q}^{(1)}
\sim {\bf \nabla}T$.  Note that the order $\varepsilon^2$ terms describe
diffusive processes, and, as assumed in Equation (8), evolve on a slower
time scale than the convective terms associated with the order $\varepsilon$
Euler equations.

To demonstrate the utility of the above LB method, we apply it to
a two-dimensional triangular lattice.  The model has one rest particle
state, $\sigma=0$, for which ${\bf e}_{\sigma i}=0$,
and two nonzero speeds for which
${\bf e}_ {\sigma i}= \sigma (\cos{(2\pi i/6)},
\sin{(2\pi i/6)})$ for $i = 0, 1\ldots,6$ and $\sigma = 1,2$.
The extension to three dimensions is straightforward and will be discussed
elsewhere~\cite{frank2}.

For this lattice geometry and the constraints discussed above,
we can solve for the coefficients of the distribution function.
One possible solution is the following:
\[ A_{0}= -\frac{5}{2}n\epsilon + n + 2n\epsilon^{2},
A_{1}=  \frac{4}{9}n\epsilon -\frac{4}{9}n\epsilon^{2},
A_{2}=  \frac{1}{9}n\epsilon^{2} -\frac{1}{36}n\epsilon,\]
\[B_{1}=   \frac{4}{9}n - \frac{4}{9}n\epsilon,
B_{2}= \frac{1}{9}n\epsilon  - \frac{1}{36}n, \]
\[ C_{1}= \frac{8}{9}n - \frac{4}{3}n\epsilon,
C_{2}=   -\frac{1}{72}n + \frac{1}{12}n\epsilon,\]
\[D_{0}=  -\frac{5}{4}n + 2n\epsilon,
D_{1}=  -\frac{2}{9}n + \frac{2}{9}n\epsilon,
 D_{2}=   \frac{1}{72}n - \frac{1}{18}n\epsilon, \]
\[E_{1}= -\frac{4}{27}n,
E_{2}= \frac{1}{108}n,
F_{1} = 0,
F_{2}=0 \]

Identifying the coefficients in Equations (1) - (3) with the corresponding
terms from the Chapman-Enskog expansion, we determine the values for the
transport coefficients.  The shear viscosity and the thermal conductivity
are given by, $\mu = n\epsilon(\tau-\frac{1}{2})$ and
$\kappa = 2n\epsilon (\tau-\frac{1}{2})$,
respectively, and yield a Prandtl number, $Pr=1/2$.
As in the case of a monatomic gas, the bulk viscosity vanishes
because $\lambda = -\mu$.

We carried out four numerical tests to determine the accuracy of
this method for simulating Equations (1)-(3). Each test focused
on one aspect of thermohydrodynamic transport.

We determined the viscosity $\mu$ by simulating an
isothermal Poiseuille flow.  Numerical results demonstate that
the model accurately reproduces a parabolic momentum profile
(not shown here). The viscosity is related to the momentum at the
channel center by $\mu = W^{2}f/8u_{cen}$ where
$f$ is the magnitude of the forcing, $u_{cen}$ is
the velocity at the center, and W is the channel width~\cite{kad}.
In Figure 1, we show the dependence of viscosity on the relaxation
time $\tau$ and two internal energies $\epsilon$.
The resulting viscosities from measurements agree
with the Chapman-Enskog theory to within around 1$\%$ over the entire range
of parameters simulated.

We determined the thermal conductivity $\kappa$ by measuring the
heat transfer across a temperature gradient, using
Fourier's law $\bf{q} = - \kappa \bf{\nabla}T$.
By fixing the temperatures at
the channel walls, we obtain a linear temperature profile and thus a
constant gradient.
Again, numerical results agree with the theoretical predictions
quite well. Since the thermal conductivity has the same functional
form as the viscosity, we also display the results in Figure 1.

For the two dimensional LB scheme, linearized perturbation
theory gives a simple relation between
the adiabatic sound speed, $c_s$ and internal energy:
$c_s = \sqrt{2\epsilon}$.
In Figure 2, we present the sound speed as a function of internal energy
for both numerical measurements and theory -- the agreement is evident.

We simulated a Couette shear flow with a temperature gradient between the
boundaries\cite{whi}.  For small temperature gradients the
pressure is essentially constant across the channel, and the
temperature profile has an analytic solution given by
$\epsilon^* = (\epsilon - \epsilon_{0})
/(\epsilon_{1} - \epsilon_{0}) = (1/2)(1+y^*)+(Br/8)(1-y^{*2})$,
where $y^*$ is the normalized distance from the center of the channel,
$\epsilon_{1}$ and $\epsilon_{0}$ are
the internal energies of the upper and lower walls, respectively.
The Brinkman number, $Br$ is the product of the Prandtl and Eckert numbers.
The agreement between theory and simulation, as shown in Figure 3,
demonstrates the validity of the method in simulating flows
in which energy dissipation is an important factor.

In conclusion, we have developed a lattice Boltzmann scheme
for the simulation of viscous, compressible, heat-conducting flows
of an ideal monatomic gas. The kinetics of this model can be easily
implemented on a parallel architecture machine. We have demonstrated
theoretically and numerically that the macroscopic behavior of this
model corresponds to that of Equations (1) - (3).
Several issues remain.
First, the current model uses the single-time-relaxation approximation,
and this restricts simulations to flows with Prandtl number $Pr= 1/2$.
In order to simulate flows with other Prandtl numbers, we should use
a full matrix collision operator, which leads to a multi-time scale
relaxation~\cite{frank2}. Second,  the equation of state in the present model
is that of ideal monatomic gas.  To simulate non-ideal gases
we may incorporate some internal degrees of freedom.
An analysis of the numerical stability of the current model
and its benchmarking against other computational fluid dynamical schemes
are under investigation.

We thank G. D. Doolen, D. W. Grunau,
B. Hasslacher, S. A. Janowsky, J. L. Lebowitz,
L. Luo, R. Mainieri and W. Matthaeus for encouragement and helpful discussions.
This work is supported by the US Department of Energy at
Los Alamos National Laboratory. Numerical
simulations were performed  on the CM-200
at the Advanced Computing Laboratory at Los Alamos National Laboratory.

\newpage
\section*{Figure Captions}

\noindent
Fig.1 : Shear viscosity ($+$) and thermal
conductivity ($\Diamond$), as functions of relaxation parameter $\tau$.
Upper curve corresponds to internal energy, $\epsilon=0.625$, lower curve
$\epsilon=0.5$.  The solid lines are the
theoretical predictions.

\noindent
Fig.2 :  Numerical simulations of adiabatic sound speed ($+$)
as a function of internal energy $\epsilon$. The solid line is the function
$\sqrt{2\epsilon}$.

\noindent
Fig.3 :  Normalized internal energy, $\epsilon^*$ for Couette flow with
heat transfer for Brinkman numbers $Br=5$ ($+$) and $Br=10$ ($\Box$).
The upper wall is moving with speed $U_{1} = 0.1$, and the lower wall
is stationary. The solid lines represent analytical results.

\end{document}